\documentclass[iop, twocolumn]{aastex63}
\usepackage{verbatim,subfigure, hyperref, threeparttable, amsmath}
\hypersetup{citecolor=blue,urlcolor=red}
\shorttitle{Radio-quiet gamma-ray pulsar}
\shortauthors{Wang et al.}

\begin{document}

\title{Searching for radio pulses from radio-quiet gamma-ray pulsars with FAST}

\correspondingauthor{SQ Wang}
\email{wangshuangqiang@xao.ac.cn}

\author{S.Q. Wang}
\affiliation{Xinjiang Astronomical Observatory, Chinese Academy of Sciences, Urumqi, Xinjiang 830011, People's Republic of China}
\affiliation{CSIRO Astronomy and Space Science, PO Box 76, Epping, NSW 1710, Australia}
\affiliation{Key Laboratory of Radio Astronomy, Chinese Academy of Sciences, Urumqi, Xinjiang, 830011, People's Republic of China}

\author{S. Dai}
\affiliation{CSIRO Astronomy and Space Science, PO Box 76, Epping, NSW 1710, Australia}

\author{N. Wang}
\affiliation{Xinjiang Astronomical Observatory, Chinese Academy of Sciences, Urumqi, Xinjiang 830011, People's Republic of China}
\affiliation{Key Laboratory of Radio Astronomy, Chinese Academy of Sciences, Urumqi, Xinjiang, 830011, People's Republic of China}

\author{A. Zic}
\affiliation{CSIRO Astronomy and Space Science, PO Box 76, Epping, NSW 1710, Australia}

\author{G. Hobbs}
\affiliation{CSIRO Astronomy and Space Science, PO Box 76, Epping, NSW 1710, Australia}

\author{D. Li}
\affiliation{New Cornerstone Science Laboratory, Department of Astronomy, Tsinghua University, Beijing 100084, People's Republic of China}
\affiliation{National Astronomical Observatories, Chinese Academy of Sciences, Beijing 100101, People's Republic of China}

\author{R.B. Ding}
\affiliation{Institute for Frontiers in Astronomy and Astrophysics, Beijing Normal University, Beijing 102206, People's Republic of China}
\affiliation{School of Physics and Astronomy, Beijing Normal University, Beijing 100875, People's Republic of China}

\author{L. Peng}
\affiliation{Key Laboratory of Stars and Interstellar Medium, Xiangtan University, Xiangtan 411105, Hunan, People's Republic of China}
\affiliation{Xinjiang Astronomical Observatory, Chinese Academy of Sciences, Urumqi, Xinjiang 830011, People's Republic of China}

\author{Z.C. Pan}
\affiliation{National Astronomical Observatories, Chinese Academy of Sciences, Beijing 100101, People's Republic of China}

\author{S.B. Zhang}
\affiliation{Purple Mountain Observatory, Chinese Academy of Sciences, Nanjing 210008, People's Republic of China}

\begin{abstract}

We present periodicity and single-pulse searches at 1250\,MHz for 22 “radio-quiet” gamma-ray pulsars, conducted using the Five-hundred-meter Aperture Spherical Radio Telescope (FAST). For PSR J1813$-$1246, we successfully detected pulsed signals with a spin period of 48.08\,ms and a dispersion measure of 209.85\,${\rm pc\,cm^{-3}}$, consistent with the spin period measured at gamma-ray wavelengths. The estimated flux density is approximately 9\,$\mu$Jy. For the remaining 21 sources, no radio emission was detected, with flux density upper limits of several $ \mu$Jy. The capability to detect pulsars with such low flux densities provides the opportunity to determine if and how these faint sources differ from much radio-brighter pulsars.

\end{abstract}


\section{INDRUCTION}

Pulsars emit broadband pulsed radiation across the electromagnetic spectrum, from radio to gamma-rays. They were first discovered at radio wavelengths in 1967 through the detection of periodic radio pulses~\citep{1968Natur.217..709H}, and a few years later, gamma-ray pulsations were observed from the Crab pulsar~\citep{1971NPhS..232...99B,1974Natur.251..397K}. 
The launch of the Large Area Telescope (LAT) aboard the \emph{Fermi} satellite has dramatically expanded the known population of gamma-ray pulsars~\citep{2010ApJS..187..460A,2013ApJS..208...17A,2023ApJ...958..191S}, with nearly 300 sources currently identified. These include both young pulsars and millisecond pulsars, as cataloged in the Public List of LAT-Detected Gamma-Ray Pulsars\footnote{\url{https://confluence.slac.stanford.edu/display/GLAMCOG/Public+List+of+LAT-Detected+Gamma-Ray+Pulsars}}.
Gamma-ray pulsars offer valuable insights into the emission mechanisms of neutron stars.

Follow-up radio observations of unassociated gamma-ray sources have led to the discovery of numerous gamma-ray pulsars. Initiatives such as the Pulsar Timing Consortium~\citep{2008A&A...492..923S} and the Pulsar Search Consortium~\citep{2012arXiv1205.3089R} have carried out targeted radio observations using major radio telescopes, including Parkes~\citep{2010PASA...27...64W}, Jodrell Bank~\citep{2004MNRAS.353.1311H}, Nançay~\citep{2011ApJ...732...47C}, Green Bank~\citep{2011ApJ...727L..16R}, Effelsberg~\citep{2013MNRAS.429.1633B}, GMRT~\citep{2013ApJ...773L..12B}, Arecibo~\citep{2016ApJ...819...34C}, Molonglo~\citep{2019MNRAS.484.3691J}, LOFAR~\citep{2017A&C....18...40B}, MeerKAT~\citep{2023MNRAS.519.5590C}, and the Five-hundred-meter Aperture Spherical Radio Telescope (FAST; \citealt{2021SCPMA..6429562W}).
Radio observations provide complementary information for gamma-ray pulsars. The measurement of dispersion measure (DM) allows distance estimates~\citep{2017ApJ...835...29Y}, which are essential for determining gamma-ray luminosities. Radio polarimetric observations yield insights into the geometry of the pulsar magnetosphere and the origin of radio emission~\citep{1969ApL.....3..225R}. The discovery of misalignment between radio and gamma-ray pulse profiles in many pulsars suggests that the two emission mechanisms are located at different regions~\citep{2010ApJ...714..810R}.

However, some gamma-ray pulsars remain undetected in the radio band~\citep{1983ApJ...272L...9B, 2010ApJ...711...64A}. Based on the distribution of flux densities among radio-loud gamma-ray pulsars, a threshold of $30\,\mu\mathrm{Jy}$ (at 1.4 GHz) is commonly used to classify pulsars as “radio-quiet” (e.g., \citealt{2023ApJ...958..191S}). Notably, “radio-quiet” does not imply the complete absence of radio emission. It may simply be too faint to detect with existing facilities. Indeed, several gamma-ray pulsars have been found to exhibit extremely faint radio emission, such as PSR J1907+0602 ($S_{1.5\,\mathrm{GHz}} = 3.4\,\mu$Jy; \citealt{2010ApJ...711...64A}), PSR J0106+4855 ($S_{0.82\,\mathrm{GHz}} = 20\,\mu$Jy; \citealt{2012ApJ...744..105P}), and PSR J0318+0253 ($S_{1.25\,\mathrm{GHz}} = 11\,\mu$Jy; \citealt{2021SCPMA..6429562W}).
Statistical studies have found that the fraction of radio-quiet gamma-ray pulsars decreases with increasing spin-down energy loss rate~\citep{2010ApJ...716L..85R,2011ApJ...727..123W,2013ApJS..208...17A}, likely due to evolving beaming geometries of radio and gamma-ray emissions. 
Understanding whether these gamma-ray pulsars without radio detections are intrinsically different from the radio pulsar population remains an open question of fundamental importance.

\begin{table*}[htp!]
\begin{center}
\caption{List of Radio-queit Gamma-Ray Pulsars}
\label{t1}
\begin{tabular}{ cccccccccc } 
 \hline
  Name & RA & DECJ & $P$ &  $\dot{E}$ &  Catalog source & Data & Integration time  &  $G$   & $T_{\rm sys}$ \\ 
   &  (deg) &  (deg) & (ms) &  $({\rm 10^{33}\, erg \,s^{-1}})$ &   &  (MJD) & (s)   &     &   \\ 
  \hline
J0357+3205&	59.47&	32.09&	444.10&	6     &	4FGL J0357.8+3204     &   60556.95   &  1040    &  16.1    & 19.8    \\
J0359+5414&	59.86&	54.25&	79.43&	1318&	4FGL J0359.4+5414 &  59915.59  & 1040  &  12.8      & 26.8  \\	
J0554+3107&	88.52&	31.13&	464.96&	56&	4FGL J0554.1+3107 &  60165.04  & 1040  &    16.0   &   19.6 \\		
J0622+3749&	95.54&	37.82&	333.21&	27   &	4FGL J0622.2+3749     &  60558.03    &   1040   &    16.1  &  19.9  \\	
J0633+1746&	98.48&	17.77&	236.97&	32   &	4FGL J0633.9+1746   & 60558.94  &  1740    &   16.3   & 24.1      \\	
J1813$-$1246&	273.35&	-12.77&	48.07&	6239&	4FGL J1813.4-1246 &  60257.33  & 1040  &  11.8     &  27.0  \\
&	&	&	&	&	 &  60394.93  &  3540 & 11.7 & 27.0  \\
&	&	&	&	&	 &  60419.86  &  4140 & 11.4 & 27.1 \\
&	&	&	&	&	 &  60783.88  &  3240 & 11.7 & 27.0 \\
J1838$-$0537&	279.73&	-5.62&	145.71&	5933&	4FGL J1838.9-0537 & 59874.44   & 1040  &    12.8   & 26.8  \\	
J1844$-$0346&	281.14&	-3.78&	112.85&	4249&	4FGL J1844.4-0345 &  59877.40  & 1040  &  15.0    &  25.9  \\	
J1846+0919&	281.61&	9.33&	225.55&	34&	4FGL J1846.3+0919 &   60182.48 & 1040  &  14.8     & 26.0  \\	
J1906+0722&	286.63&	7.38&	111.48&	1022&	4FGL J1906.4+0723 &  59924.36  & 1040  &  13.5    & 26.7   \\	
J1932+1916&	293.08&	19.28&	208.21&	407&	4FGL J1932.3+1916 & 60165.70   & 1040  &  16.2    &  22.7  \\	
J1954+2836&	298.58&	28.60&	92.71&	1048&	4FGL J1954.3+2836 &  59915.43  & 1040  &   14.8    & 26.0   \\	
J1957+5033&	299.41&	50.56&	374.81&	5&	4FGL J1957.6+5033 &   60536.69 & 1040  &  14.4    &  26.3  \\	
J1958+2846&	299.67&	28.76&	290.39&	342&	4FGL J1958.7+2846 &  60165.73  & 1040  &    16.2  & 22.5   \\	
J2017+3625&	304.48&	36.42&	166.75&	12&	4FGL J2017.9+3625 &  60536.72  & 1040  &   16.2   & 21.6   \\	
J2021+4026&	305.38&	40.45&	265.32&	114 &	4FGL J2021.5+4026     &   60566.64   &   1040   &   16.2   &  23.2  \\	
J2028+3332&	307.08&	33.53&	176.71&	35&	4FGL J2028.3+3331 &  60164.78  & 1040  &   13.5    &  26.7 \\	
J2030+4415&	307.72&	44.26&	227.07&	22&	4FGL J2030.9+4416 & 60536.74   & 1040  &  15.2    & 25.7   \\
J2055+2539&	313.95&	25.67&	319.56&	5&	4FGL J2055.8+2540 &  60535.76  & 1040  &  15.9    &  24.7  \\	
J2111+4606&	317.85&	46.11&	157.83&	1436&	4FGL J2111.4+4606 &   59914.44 & 1040  &  16.2    & 22.6   \\	
J2139+4716&	324.98&	47.27&	282.85&	3&	4FGL J2140.0+4716 &  60535.79  & 1040  &    13.7   & 26.6  \\	
J2238+5903&	339.62&	59.06&	162.73&	888&	4FGL J2238.5+5903 &  60164.81  & 1040  &    13.2  &   26.7 \\
\hline
\end{tabular}
\end{center}
\end{table*}

{FAST offers high sensitivity, capable of detecting flux densities down to a few $\mu$Jy~\citep{2011IJMPD..20..989N,2018IMMag..19..112L,2020RAA....20...64J,2021RAA....21..107H},  }which is significantly below the typical threshold of 30\,$\mu$Jy used to classify gamma-ray pulsars as “radio-quiet.” This makes FAST an ideal instrument for searching for faint radio emissions from gamma-ray pulsars. 
In this paper, we present periodicity and single-pulse searches using FAST for 22 gamma-ray pulsars that have remained undetected at radio wavelengths. 
The observations and data processing are described in Section 2. The results of the searches are presented in Section 3, and the discussion and summary are provided in Section 4.

\section{OBSERVATIONS AND DATA PROCESSING}

\subsection{Observations}

There are approximately 28 gamma-ray pulsars without radio detections within FAST's sky coverage. Six of them have already been studied with FAST, including PSRs J0633+0632, J1653$-$0158, J1826$-$1256, J1827$-$0849, J1836+5925, and J2034+3632 (Ding et al., in prep.). We observed the remaining 22 pulsars using the central beam of the 19-beam receiver on FAST, covering a frequency range of 1.05--1.45,GHz \citep{2010ApJ...711...64A,2020RAA....20...64J}.
The data were recorded in search mode in PSRFITS format, with four polarizations, 8-bit resolution, 4096 frequency channels, and a sampling interval of 49.152\,$\mu$s.  
 For PSR J1813$-$1246, four observations were conducted with durations of 1040\,s, 3540\,s, 4140\,s and 3240\,s, respectively.  PSR J0633+1746 was observed once, with an integration time of 1740\,s.
The remaining 20 pulsars were each observed once, with an integration time of 1040\,s.  
The total observing time amounted to approximately 9.6 hours.  
Further details of the observations are provided in Table~\ref{t1}.

\begin{figure*}
\centering
\includegraphics[width=90mm]{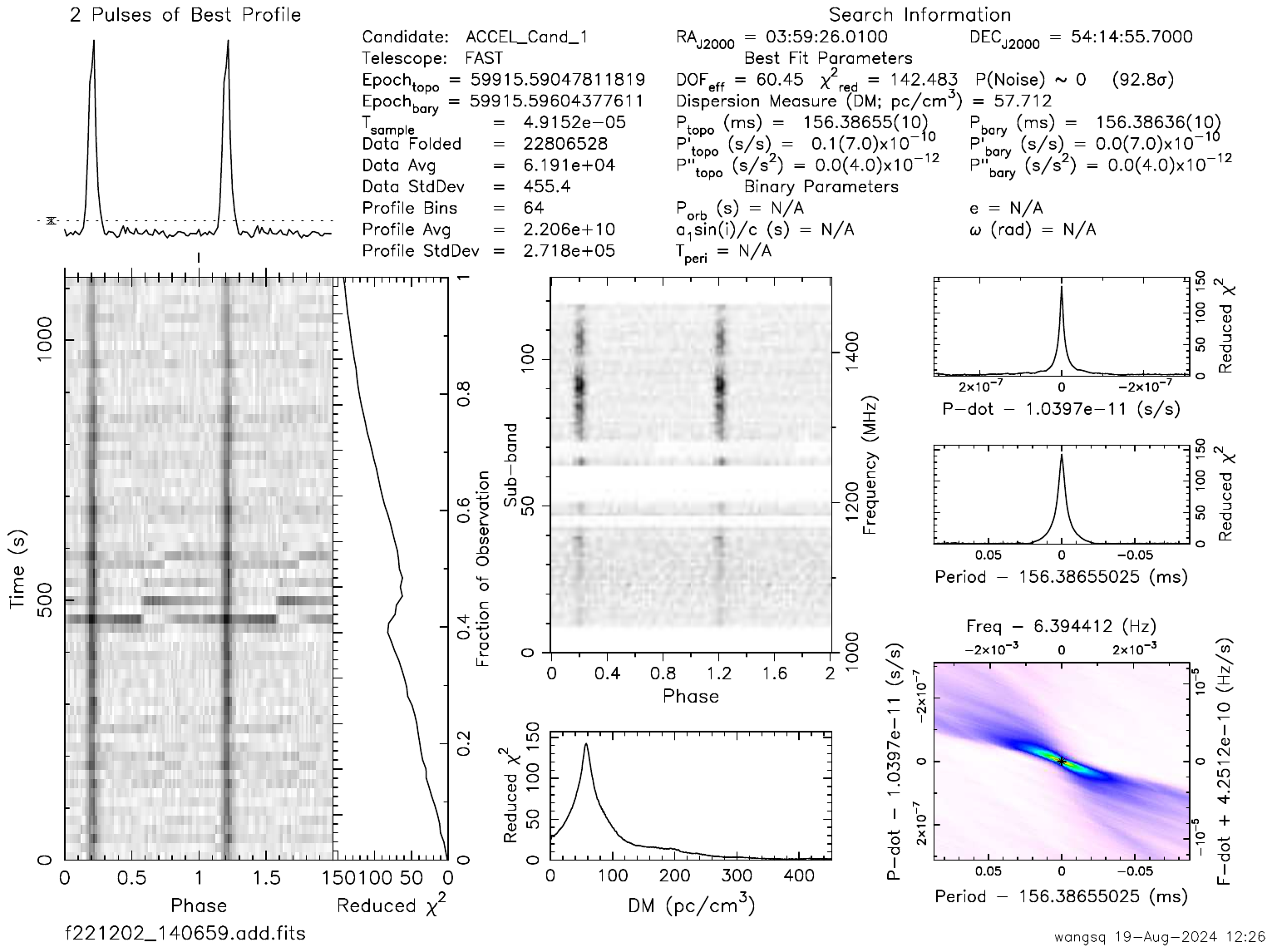}
\includegraphics[width=63mm]{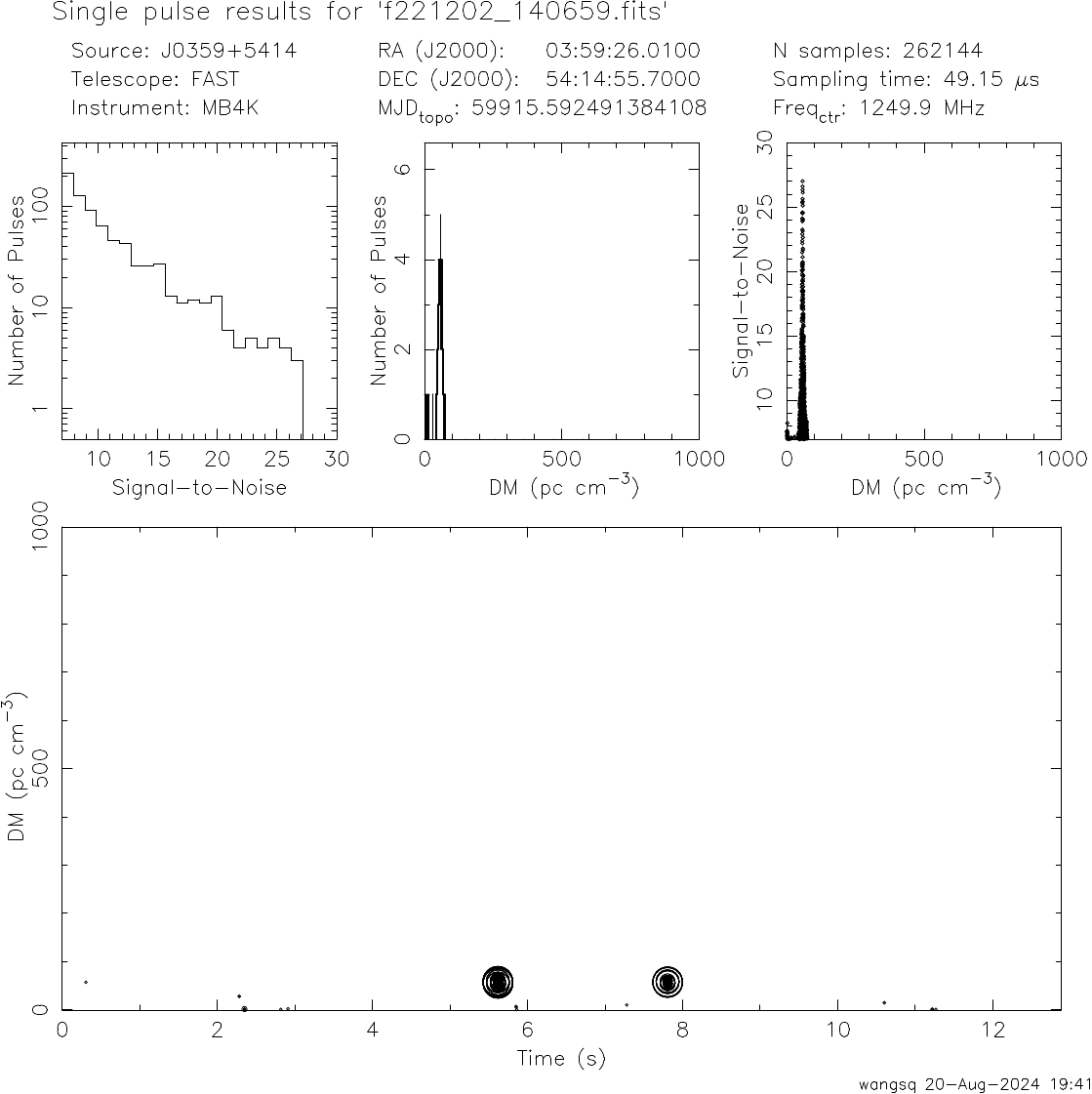}
\caption{Left panel: Periodicity search detection of PSR J0358+5413, showing a clear detection with a spin period of 156\,ms and a DM of 57.712\,${\rm pc\,cm^{-3}}$. Right panel: Single-pulse search detection of PSR J0358+5413. Several bright single pulses are detected at approximately 5.7\,s and 7.8\,s, with a consistent DM of about 57.7\,${\rm pc\,cm^{-3}}$.}
\label{0359p}
\end{figure*}

\begin{figure*}
\centering
\includegraphics[width=80mm]{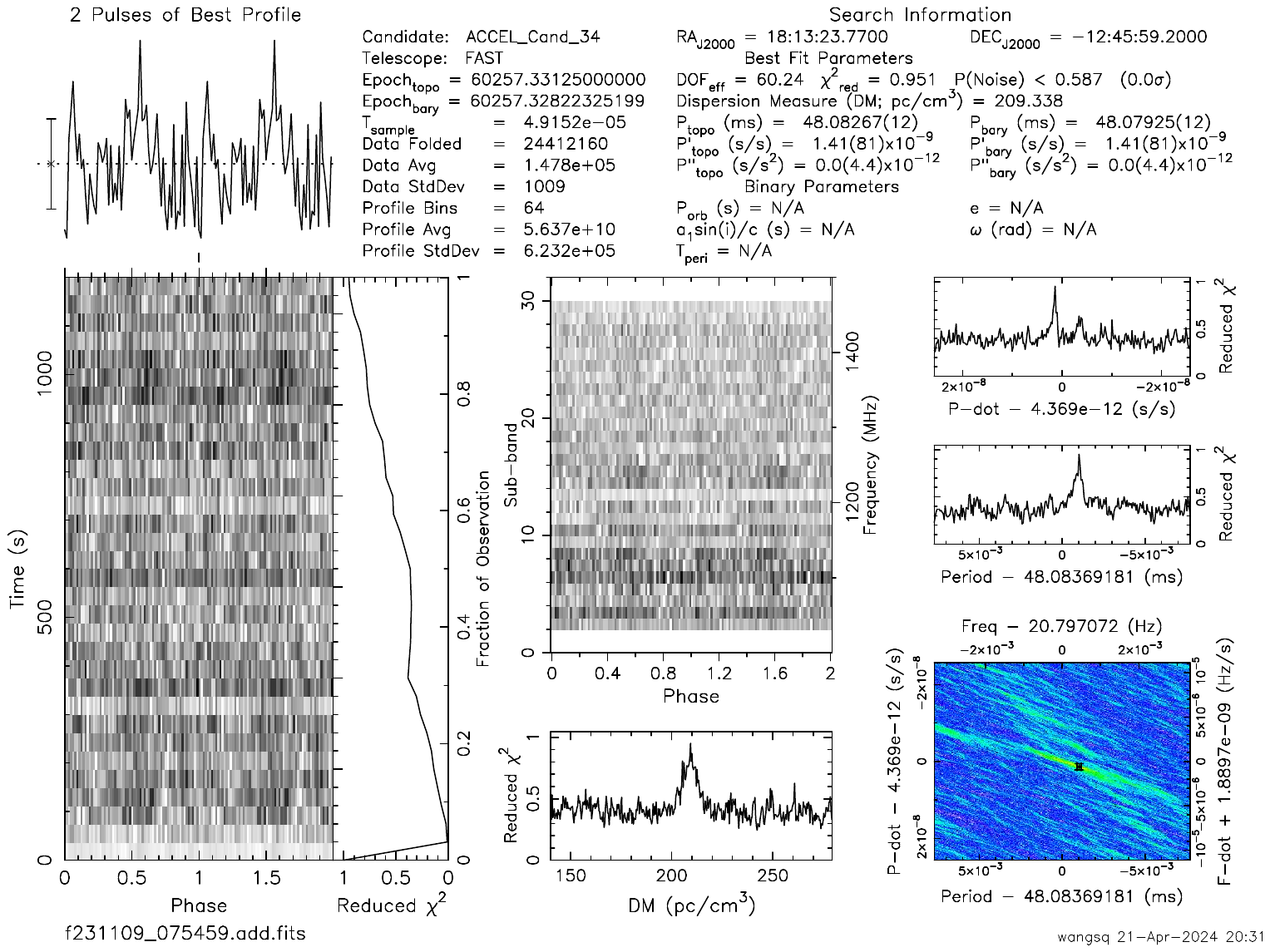}
\includegraphics[width=80mm]{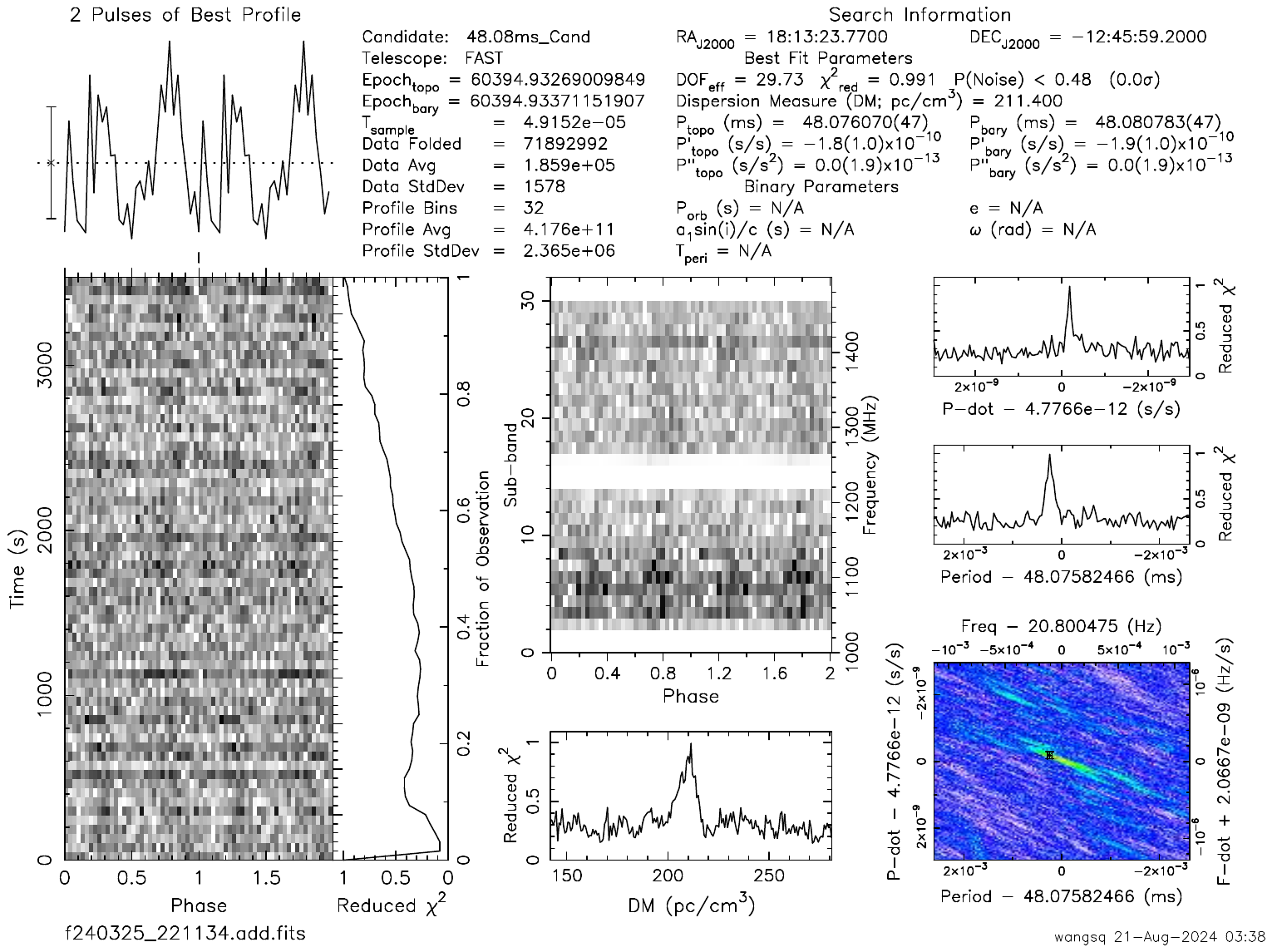}
\includegraphics[width=80mm]{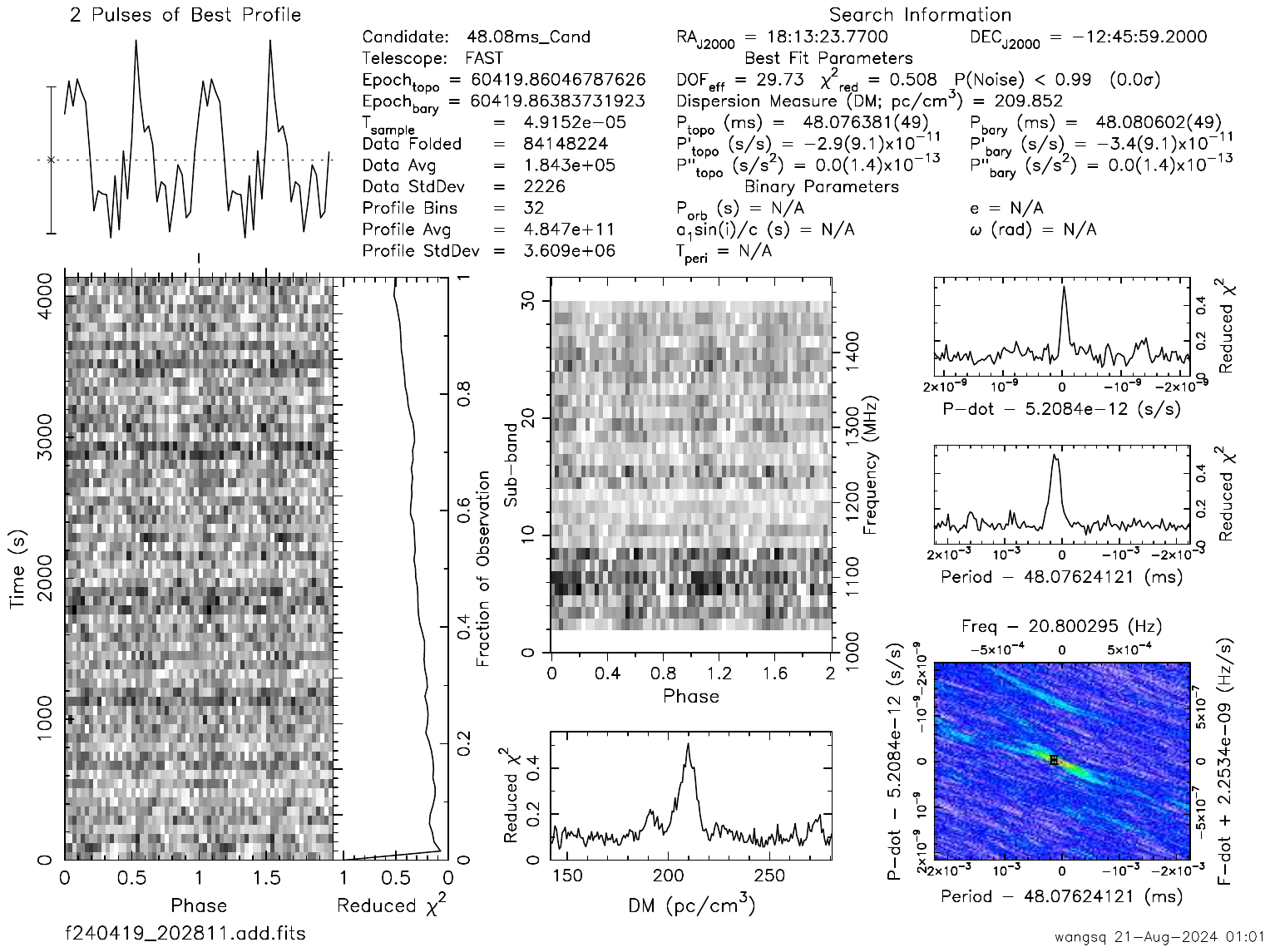}
\includegraphics[width=80mm]{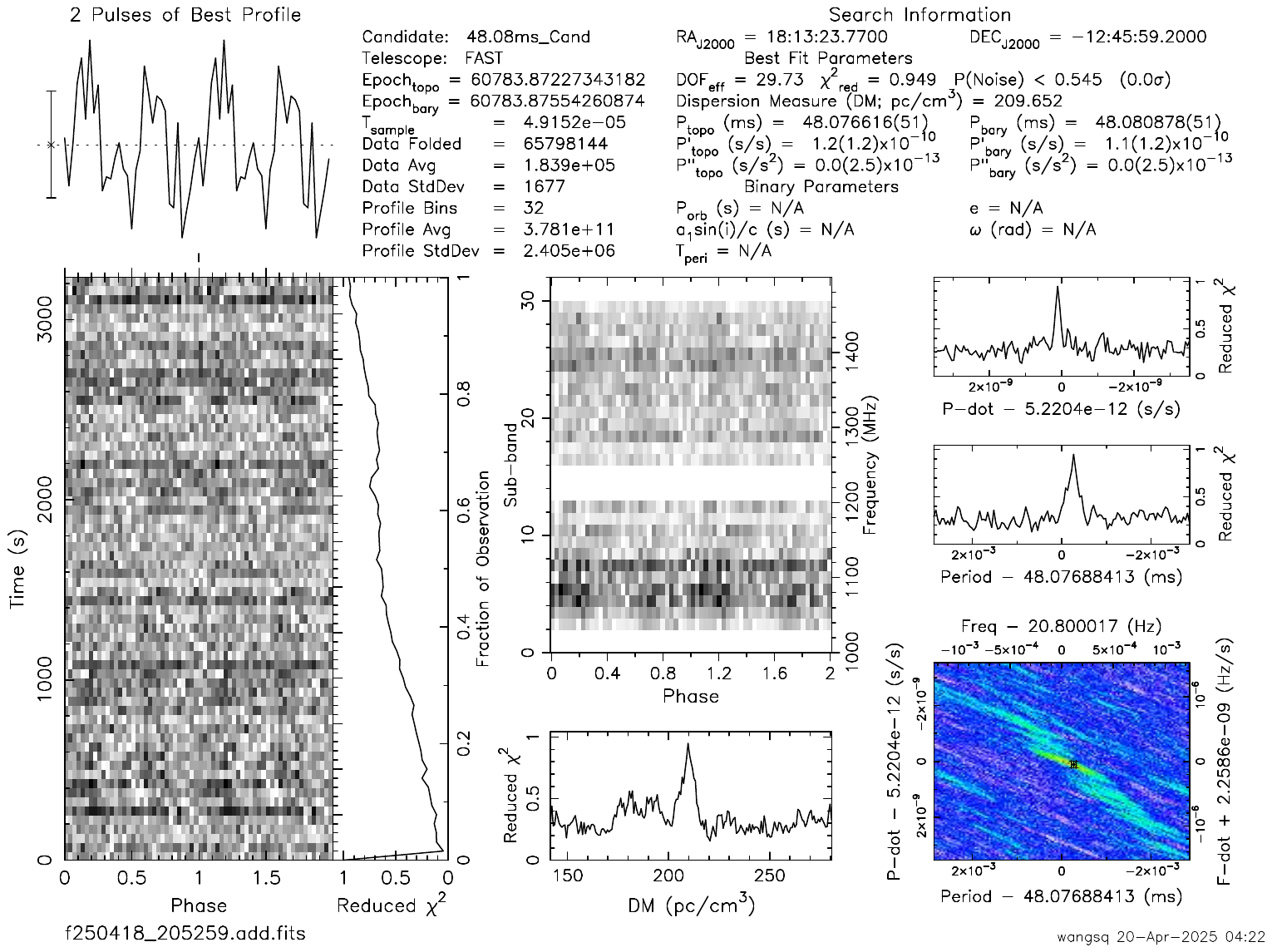}
\caption{Periodicity search detections of PSR J1813$-$1246 for observations conducted on 11 September 2023, 25 March 2024, and 20 April 2024, and 19 April 2025, with durations of 1040\,s, 3540\,s, 4140\,s, and 3240\,s, respectively.}
\label{1813p}
\end{figure*}

\subsection{Periodicity search}

We performed periodicity searches for each pulsar using {\sc PRESTO}~\citep{2001PhDT.......123R,2002AJ....124.1788R}. 
The script {\sc DDplan.py} was employed to determine the optimal strategies for de-dispersing the data. 
The data were de-dispersed with dispersion measures (DM) ranging from 0 to 1000\,${\rm pc \,cm^{-3}}$, using DM steps of 0.05, 0.1, and 0.2 ${\rm pc \, cm^{-3}}$ for the DM ranges $0$--$379$, $379$--$682$, and $682$--$1000$\,${\rm pc \,cm^{-3}}$, respectively. 
Narrowband and/or short-duration broadband radio frequency interference (RFI) was identified using {\sc rfifind}, and approximately 16\% of the frequency channels were excised. A detailed characterization of the RFI environment at FAST is provided in \citet{2020RAA....20...64J}.
As all pulsars in our sample are isolated, a zmax value of 0 was used in the acceleration search. 
We used {\sc accel\_sift.py} to reject bad candidates and consolidate multiple detections of the same candidate signals across different DMs. 
For candidates identified by {\sc accel\_sift.py}, we folded the raw data files using {\sc prepfold} according to the DM and period values. 
Typically, each observation yields approximately 500 candidates, which are then inspected manually.

Since all pulsars in our sample have been well timed through gamma-ray observations, we also used DSPSR~\citep{Straten2011} to extract single pulses based on the timing ephemerides provided by the pulsar catalog (PSRCAT; \citealt{2005AJ....129.1993M}), using an arbitrarily selected DM of 100\,${\rm pc\,cm^{-3}}$.
Subsequently, we employed {\sc pdmp} in {\sc PSRCHIVE}~\citep{2004PASA...21..302H} to search for pulsed signals for each pulsar, exploring a DM range from 0 to 1000\,${\rm pc\,cm^{-3}}$.

\subsection{Single-pulse search}

Some pulsars may emit sporadic single-pulses that are difficult to detect using periodicity searches, as seen in the case of rotating radio transients~\citep{2006Natur.439..817M}.
Motivated by this, we conducted single-pulse searches for each pulsar in our sample.  
We used {\sc PRESTO} for the single-pulse search, applying trial DMs ranging from 0 to 1000\,${\rm pc \,cm^{-3}}$.  
Candidates with a signal-to-noise ratio (S/N) greater than 7 were identified using {\sc single\_pulse\_search.py} on each de-dispersed time series, yielding approximately 6000 candidates per observation.
For each candidate, we folded the raw data using {\sc DSPSR} and performed a visual inspection of the results.

To test our pipeline, we note that the known radio pulsar J0358+5413 is located near the radio-quiet gamma-ray pulsar J0359+5414, which is included in our sample.  
The beam size of the central beam of the 19-beam receiver is approximately 3 arcminutes in the L-band, which is sufficient to cover both pulsars.  
The left panel of Figure~\ref{0359p} presents the results of the periodicity search for PSR J0358+5413 using our pipeline.  
This pulsar is clearly detected, with a spin period of 156\,ms and a DM of 57.712\,${\rm pc \,cm^{-3}}$, consistent with published results~\citep{2023RAA....23j4002W}.  
We also performed a single-pulse search for PSR J0358+5413.  
A segment of the data with a duration of 12.885\,s is shown in the right panel of Figure~\ref{0359p}, where bright single-pulses are detected at approximately 5.7\,s and 7.8\,s, with a consistent DM of approximately 57.7\,${\rm pc \,cm^{-3}}$.

\section{RESULTS}

\subsection{Periodicity search sensitivity}

The minimum detectable flux density for periodicity searches is given by \citet{2004hpa..book.....L}:
\begin{equation}
\label{eq1}
S_{\min} = \frac{(S/N)\beta T_{\rm sys}}{G(n_{\rm p}t_{\rm int}\Delta F)^{1/2}} \left(\frac{W}{P-W}\right)^{1/2},
\end{equation}
where \(G\) is the telescope gain, \(T_{\rm sys}\) is the system temperature, \(n_{\rm p}\) is the number of polarizations, \(\Delta F\) is the observational bandwidth, \(\beta\) is the sensitivity degradation factor, \(W\) is the pulse width, \(P\) is the spin period, and \(S/N\) is the threshold signal-to-noise ratio required for detection. 

For FAST, both \(G\) and \(T_{\rm sys}\) vary with zenith angle~\citep{2020RAA....20...64J}. For our observations, the values of \(G\) and \(T_{\rm sys}\) are listed in Columns 9 and 10 of Table~\ref{t1}, respectively. 
The observed pulse width \(W\) is influenced by intrinsic pulse width, interstellar scattering, intrachannel dispersive smearing, dedispersion errors, and receiver filter response time~\citep{2003ApJ...596.1142C}. In our case, dedispersion errors and filter response times are negligible, so we consider only smearing and scattering effects. 

The intrachannel dispersive smearing delay is given by:
\begin{equation}
\label{smearing}
\Delta t =  8.3 \, \mu\text{s} \cdot \text{DM} \cdot \Delta \nu_{\rm MHz} \cdot \nu_{\rm GHz}^{-3},  
\end{equation}
where \(\Delta \nu\) is the channel bandwidth. For our observations, \(\Delta \nu = 0.122\,\text{MHz}\) and \(\nu = 1250\,\text{MHz}\). For a pulsar with \(\text{DM} = 1000 \,\rm pc cm^{-3} \), the resulting \(\Delta t\) is approximately 0.5\,ms. Given that spin periods in our sample range from tens to hundreds of milliseconds, this smearing is negligible.

The expected scattering timescale is estimated as:
\begin{equation}
\label{scattering}
\begin{aligned}
\log \tau_{\rm d} = -3.72 + 0.411 \, \log \, \text{DM} + 0.937 (\log \, \text{DM})^2 \\
- 4.4 \, \log \, \nu_{\rm GHz} \quad \mu\text{s}.
\end{aligned}
\end{equation}
For example, a pulsar with \(\text{DM} = 500\, \rm pc \,cm^{-3} \) has \(\tau_{\rm d} \approx 7\,\text{ms}\), indicating that scattering significantly affects the minimum detectable flux density, particularly for high-DM pulsars.

Among the 22 sources in our sample, no radio pulsations were detected for 21 pulsars, while pulsed signals were successfully detected from PSR J1813$-$1246. 
For the 21 non-detection sources, we conservatively assumed a 10\% duty cycle and adopted a detection threshold of \(S/N = 7\), \(n_{\rm p}=2\), \(\beta=1\), \(\Delta F = 400\,\text{MHz}\), and $t_{\rm obs}=1040/1740\,s$ (Column 8 of Table~\ref{t1}). The corresponding minimum detectable flux densities are listed in the penultimate column of Table~\ref{t2}. 
Note that broad baseline features, caused by red noise, can significantly degrade sensitivity, especially for pulsars with periods longer than 100\,ms~\citep{2015ApJ...812...81L}. As a result, our estimated upper limits may underestimate the true sensitivity limits.

\subsubsection{PSR J1813$-$1246}

\begin{figure}
\centering
\includegraphics[width=80mm]{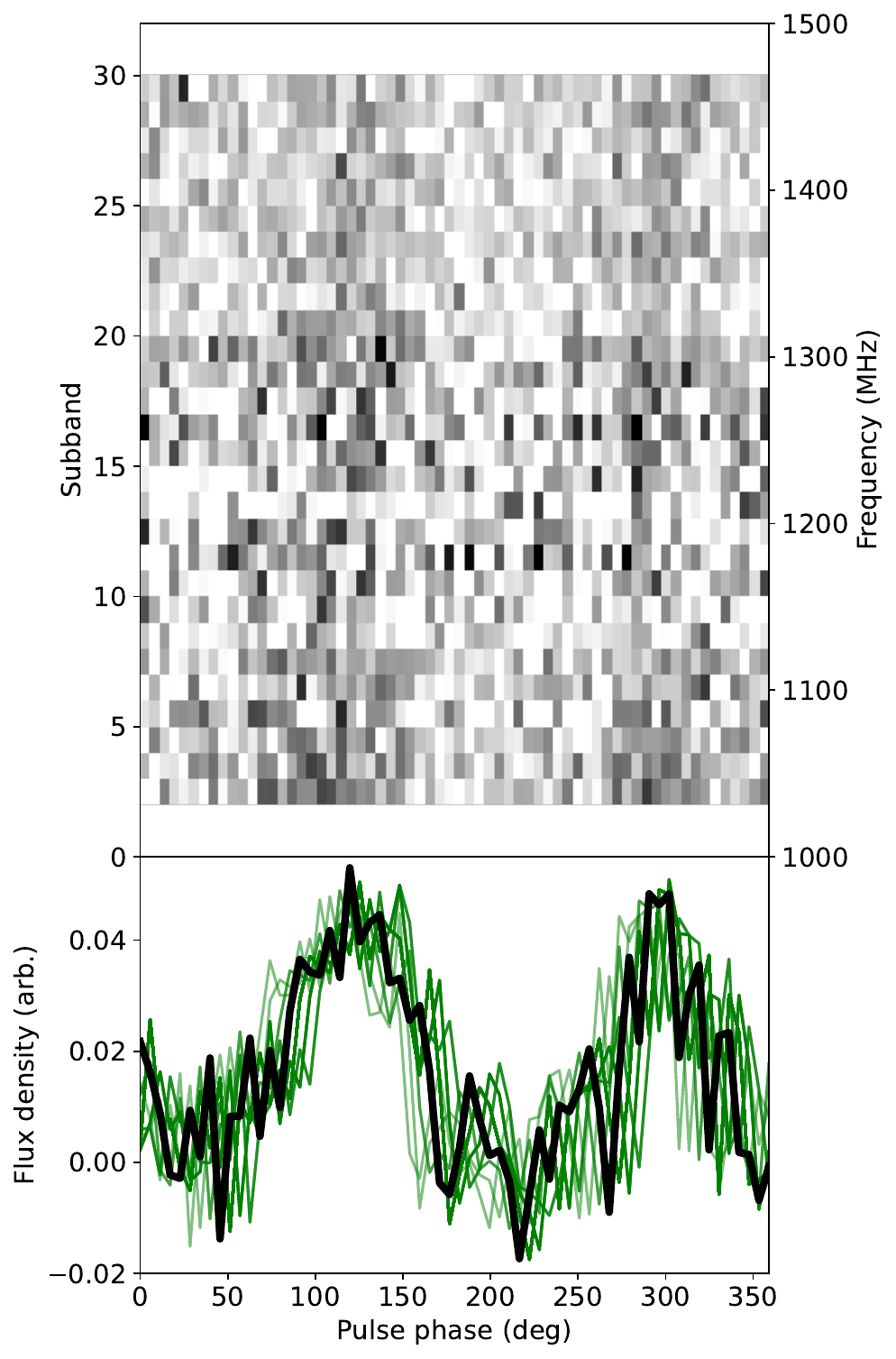}
\caption{The dynamic spectrum (top panel) and the averaged pulse profile obtained by combining all four observations (bottom panel) for PSR J1813$-$1246. The black solid line represents the final averaged profile, while the green lines show 100 averaged profiles generated through Monte Carlo simulations to estimate the uncertainty.}
\label{1813}
\end{figure}

\begin{table*}[htp!]
\caption{List of flux density upper limits for 22 radio-quiet gamma-ray pulsars.}
\label{t2}
\begin{center}
\begin{tabular}{ccccccccccccccccc} 
\hline
Name &  $S_{34}$  & $S_{150}$ & $S_{318}$ & $S_{327}$ & $S_{350}$ & $S_{430}$ &  $S_{820}$ & $S_{1398}$ &  $S_{1400}$  &  $S_{1510}$ & $S_{1520}$ &  $S_{2000}$ & $S_{1250}$  & Ref. \\ 
\hline
J0357+3205      &   34      &  0.8   &     -            &  0.043   &   0.134   &   -       &   -         &     -       &  -                &    -           &    -       &     -        &   0.003    &   2,6,9  \\
J0359+5414 	 &   -          &  1.7  &     -       	  	 &    -      &    -       &     -      &    -         &   -      &	0.015           &   -            &    -      &   0.018  & 0.005      &  8,9    \\ 	
J0554+3107 	 &   -  	 & 1.2  &     -        	 &     -       &        -     &     -      &     -      &     -        &	0.066   &  -             &   -       &    -         &     0.003  &   5,9  \\
J0622+3749      &    -         & 1.4   &     -       	 &     -       &  0.131    &    -       &  0.032  &  -          &	0.022   &    -           &     -     &    -         &     0.003 &   3,9  \\
J0633+1746      &   38 	  &  -   &     0.13             &    -        &  -           &        0.032    &     -     &       -   &     0.025         &      -       &      -            &           -        &     0.003  &   6,10  \\
J1813$-$1246       &   1179     &  -    	 &    -    &          -      &   -           &      -     &  0.042 &  -         &      -           &      -       &   -        &    0.028 &  0.009     &  2,6  \\
J1838$-$0537       &    -          & 26    &     -       	 &     -       &   -           &       -    & 0.082  &  -         &      -            &     -        &   -        &   0.009  &  0.005    &   4,9    \\   	
J1844$-$0346       &    -          &   -  	&     -        &      -      &   -           &       -    &       -   &    -    &	0.021    &    -         &   -        & 0.024    &  0.004    &    8    \\   	
J1846+0919 	 &    774      & 7.6  &     -       	 &      -      & 0.209      &     -    &       -   &  -        &      -             & 0.004    &  -         &  -            &  0.004    &   1,6,9   \\	
J1906+0722 	 &    -          & 8.0  	&     -        &      -      &   -           &     -     &        -   &   -          &      0.021     &      -     &   -         &    -         &  0.005    &    7,9   \\	
J1932+1916 	 &    -         & 2.9   &     -       	 &     -        &  -           &      -     &       -   &  -       &      0.075      &     -        &  -         &   -          &   0.004 &  5,9     \\	
J1954+2836      &    614      & 2.1   &     -       	 &     -        &  -            &     -     &      -    &   -      &      -             & 0.004    &  -         &  -           &  0.004   &   1,6,9    \\	
J1957+5033 	 &    527     &  1.3    &     -       	 &    -         &  0.122    &           &  0.025   &  -     &      -             &      -      &   -        &   -           &   0.005  &    1,6,9    \\   	
J1958+2846 	 &    642     &  2.0   &     -       	 &     -        &  -           &      -     &       -    &     -     &      -    &  0.005  &  -         &  -            &    0.004  &  2,6,9   \\ 	
J2017+3625 	 &    -         &  1.3   &     -       	 &     0.113 &  -          &     -     &   0.043 & 0.050    &	0.017     &   0.005 &  -         &   0.010   &  0.003   &    8,9    \\
J2021+4026      &    92	  &  3.2   &     -       	 &     -        &  -           &     -     &    0.051 &  -     &      -             &   -        &   -         &     0.011  &   0.004  &   2,6,9    \\
J2028+3332 	 &    -          &  1.9   &     -       	 &   0.142  &  -           &     -     &    0.033 &  -    &      -             &  0.004  &    -        &  0.015     &   0.005  &   9    \\ 	
J2030+4415 	 &    -          &  2.2  &     -       	 &      -        &   -          &    -     &    0.019 &  -    &       0.023    &   -         &   -         &    -         &    0.004  &   3,9   \\              	
J2055+2539 	 &    60      &  1.3  	&     -        &  0.085    &  0.110    &    -     &        -    &    -      &       -    &   0.106 &  -        &  -            &     0.004 &   1,6,9   \\      	
J2111+4606 	 &    -          &  1.9   &     -       	 &     -        &   -           &     -     &  0.033  &  -        &       -     &      -      & 0.014 &   -            &   0.004  &    3,9   \\   	
J2139+4716 	 &    74        &  1.1  &     -        	 &     -        &   0.171    &    -     &   0.034 &  -     &      0.022      &   -         &    -      &  -            &   0.005  &    3,6,9   \\     	
J2238+5903 	 &    82       &  1.9   &     -       	 &     -        &   -           &     -     &   0.027 &  -    &     -             &      -      &     -     &  0.007     &  0.005   &    2,6,9   \\
\hline
\end{tabular}
\end{center}
\footnotesize{Column (1) presents pulsar name. Column (2) to (13) present the flux density  upper limits with the units of mJy at 34, 150, 327, 350, 430, 820, 1398, 1400, 1510, 1520, 2000, and 1250\,MHz, respectively.  Column (14) provides references: 1. \citet{2010ApJ...725..571S}, 2. \citet{2011ApJS..194...17R}, 3.  \citet{2012ApJ...744..105P}, 4. \citet{2012ApJ...755L..20P}, 5. \citet{2013ApJ...779L..11P}, 6. \citet{2014MNRAS.445.3221M}, 7. \citet{2015ApJ...809L...2C}, 8. \citet{2018ApJ...854...99W}, 9. \citet{2021A&A...654A..43G}, 10 \citet{2000AIPC..513..353B}.  }

\end{table*}

\begin{figure*}
\centering
\includegraphics[width=150mm]{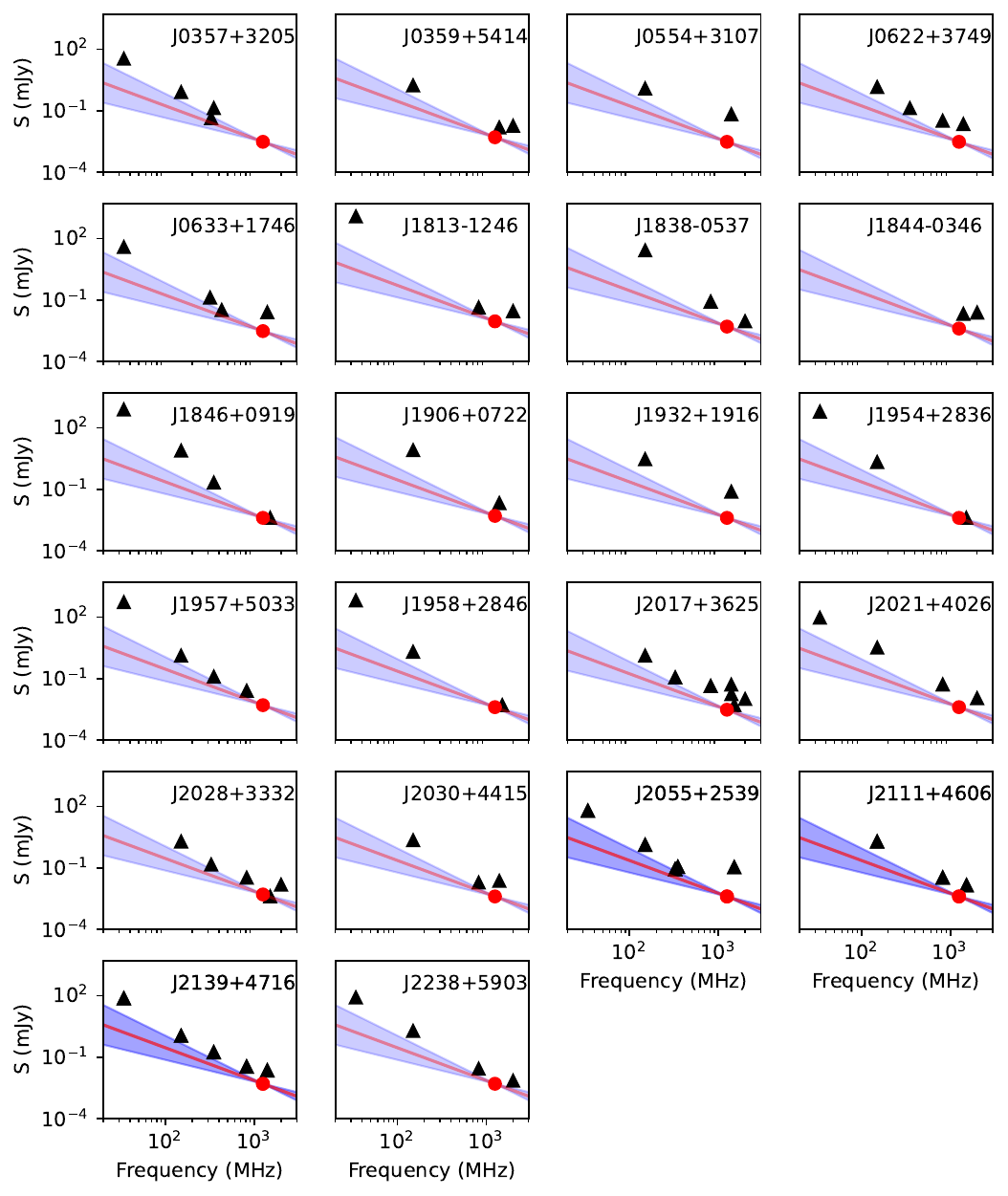}
\caption{Flux density upper limits for 22 radio-quiet gamma-ray pulsars. The red circles/stars represent the flux density upper limits/detections from our observations at 1250\,MHz, while the black triangles denote results from previous observations. More details are provided in Table~\ref{t2}. The blue shaded region indicates the range of equivalent flux density limits from our observations, assuming a spectral index of $-1.60 \pm 0.54$~\citep{2018MNRAS.473.4436J}. The red line represents the expected flux density trend for a spectral index of $-1.60$.}
\label{flux}
\end{figure*}

PSR J1813$-$1246 is a radio-quiet gamma-ray pulsar with two distinct components in its gamma-ray profile~\citep{2013ApJS..208...17A}. We conducted a radio observation of this pulsar using FAST on 11 September 2023, with a duration of 1040\,s. A very weak signal was detected during this observation (Figure~\ref{1813p}), but the S/N was insufficient to confirm a detection. The observed spin period, approximately 48.08\,ms, closely matches the period measured at gamma-ray wavelengths.

To improve detection significance, we conducted three additional observations, constrained by the maximum tracking time of FAST. These were carried out on 25 March 2024, 20 April 2024, and 19 April 2025, with integration times of 3540\,s, 4140\,s, and 3240\,s, respectively. Pulses with similar periods were detected in all three observations, although the signals remained weak (Figure~\ref{1813p}). Across the four observations, the signals consistently exhibited DMs of 209.33, 211.40, 209.85, and 209.65\,${\rm pc \,cm^{-3}}$, respectively. Based on these consistent detections, we confirm the presence of faint radio emission from PSR J1813$-$1246.

The radio profile shows a main pulse and an interpulse, both with a similar W50 width of $\sim$0.2\,P, comparable to the gamma-ray profile, but with a narrower pulse width. Taking the longest observation of 4140\,s, we calculated an S/N of approximately 8 and a duty cycle of 0.4 using the {\sc PSRCHIVE} software~\citep{2004PASA...21..302H}. We obtained an estimated flux density of approximately 9\,$\mu$Jy.

Using the DM value of 209.85\,${\rm pc\,cm^{-3}}$, we refolded each observation based on the gamma-ray ephemeris from \citet{2011ApJS..194...17R} to obtain individual profiles. An analytic template was constructed from the high S/N observation on 20 April 2024 using {\sc paas} tool in the {\sc PSRCHIVE} software package. We then used {\sc pat} to measure the phase offsets of the remaining three observations relative to this template. After aligning the profiles in phase, we summed them to produce the averaged pulse profile, shown as the black line in Figure~\ref{1813}.
To estimate the uncertainty of the averaged profile, we performed a Monte Carlo simulation. In this analysis, we applied random phase shifts drawn from Gaussian distributions, with standard deviations corresponding to the uncertainties reported by {\sc pat} for each observation. We repeated this process 100 times, and the resulting averaged profiles are shown as green lines in Figure~\ref{1813}.

The detection of radio emission allows us to estimate the distance to PSR J1813$-$1246. Using the DM value from the longest observation of 209.852\,${\rm pc \, cm^{-3}}$, the estimated distances are 4.25\,kpc and 3.97\,kpc, based on the YMW16~\citep{2017ApJ...835...29Y} and NE2001~\citep{2002astro.ph..7156C} electron-density models, respectively. The energy flux of this pulsar at gamma-ray wavelengths is $25.3 \times 10^{-11} \, \rm erg\,cm^{-2}\,s^{-1}$~\citep{2013ApJS..208...17A}. The gamma-ray luminosity can be calculated using the formula:
\begin{equation}
\label{gamma}
L_{\gamma}=4 \pi d^2 f_{\Omega} G_{100},
\end{equation}
where \(d\) is the distance, \(f_{\Omega}\) is the beam correction factor, and \(G_{100}\) is the integrated energy flux in the 0.1 to 100 GeV energy band. Assuming \(d = 4.25\,\rm kpc\) and \(f_{\Omega} = 1\), we obtain \(L_{\gamma} \sim 5 \times 10^{35}\,{\rm  erg \,s^{-1} }\), and the gamma-ray efficiency (\(L_{\gamma}/\dot{E}\)) is approximately 8\%, which is consistent with values observed in radio-loud gamma-ray pulsars~\citep{2013ApJS..208...17A}.

\subsection{Single-pulse search sensitivity}

For single-pulse searches, the minimum detectable peak flux density is given by~\citet{2003ApJ...596.1142C}:
\begin{equation}
\label{eq2}
S_{\min} = \frac{(S/N_{\rm peak}) 2 \beta T_{\rm sys}}{G(n_{\rm p} W \Delta F)^{1/2}},
\end{equation}
where \(S/N_{\rm peak}\) is the peak signal-to-noise ratio of a pulse.

In our study, no significant single-pulses were detected for any of the 22 pulsars. Assuming a minimum $S/N_{\rm peak} = 7$ and an arbitrary pulse width of 20\,ms, we estimate that the minimum detectable flux density for a single pulse is on the order of several mJy. However, pulse widths can vary significantly between single pulses for different pulsars (e.g.,~\citealt{2010MNRAS.402..855B,2020ApJ...902L..13W,2024ApJ...964....6W}), resulting in large uncertainties in the upper-limit flux density estimations.

\section{ DISCUSSION AND CONCLUSIONS}

We conducted periodicity and single-pulse searches for 22 radio-quiet gamma-ray pulsars. For 21 of these sources, no radio emission was detected. Stringent upper limits on the flux densities at 1250\,MHz were obtained, typically at the level of several $\mu$Jy (see the penultimate column of Table~\ref{t2}). In contrast, pulsed radio signals were detected from PSR J1813$-$1246, with an estimated flux density of $\sim9\,\mu$Jy. Flux density upper limits from previous observations for these sources are also summarized in Table~\ref{t2}.
Pulsars exhibit diverse spectral properties. For most sources, their radio spectra can be described by a simple power-law model, although some pulsars exhibit spectral flattening or turnover at low frequencies~\citep{1995MNRAS.273..411L}. In our analysis, we adopted a power-law spectral model with a mean spectral index of $-1.60$ and a standard deviation of 0.54~\citep{2018MNRAS.473.4436J}, illustrated by the shaded regions in Figure~\ref{flux}. Our flux density upper limits (red circles), detection (red star), and previous measurements (black triangles) are plotted in Figure~\ref{flux}. The upper limits derived from our FAST observations are generally more stringent than those obtained in previous studies at other frequencies. We note, however, that this comparison is approximate due to uncertainties in the individual spectral indices of the pulsars.

The physical mechanisms responsible for the apparent radio quietness of gamma-ray pulsars remain uncertain. Several explanations have been proposed for the absence of detectable radio emission, including intrinsic weakness of the radio signal or a radio beam that does not sweep across the Earth. If the non-detections are primarily due to sensitivity limitations, then observations with highly sensitive radio telescopes may reveal weak radio signals. Indeed, follow-up observations using sensitive instruments have detected faint radio emissions from a few gamma-ray pulsars, including PSR J1813$-$1246 in our sample, as well as PSR J1907+0602~\citep{2010ApJ...711...64A} and PSR J0318+0253~\citep{2021SCPMA..6429562W}. Recently, \citet{2024ATel16875....1D} reported a detection of PSR J0359+5414 with a 6600\,s integration using FAST, while our observation of the same pulsar with an integration time of 1040\,s yielded no detection. Notably, PSRs J1813$-$1246, J0359+5414, and J1907+0602 all exhibit large spin-down energy loss rates with $\dot{E} > 10^{36}$\,erg\,s$^{-1}$, which may further support that the fraction of radio-quiet pulsars decreases with increasing $\dot{E}$~\citep{2010ApJ...716L..85R}.

The discovery of lags between radio and gamma-ray pulse profiles suggests that these emissions originate from different regions within the pulsar magnetosphere~\citep{2013ApJS..208...17A}. 
Radio emission is believed to originate from the polar cap region~\citep{1975ApJ...196...51R}, forming a relatively narrow beam compared to gamma-ray emission~(e.g., \citealt{1988MNRAS.234..477L}). The morphology of the radio beam is complex, with models including the core-cone model~\citep{1983ApJ...274..333R} and the patchy model~\citep{1988MNRAS.234..477L}.  
Detailed two-dimensional beam mapping of PSR J1906+0746 revealed that the beam may fill the open field line region latitudinally but is less extended longitudinally, indicating a non-canonical hollow-cone structure~\citep{2019Sci...365.1013D}. Although the exact beam shape remains uncertain, it is now recognized that radio emission beams are generally underfilled in longitude, with only certain regions actively emitting~\citep{2019MNRAS.490.4565J, 2025ApJ...983..179S}. As a result, the observed radio profile may appear much narrower than predicted by idealized models, or even remain undetected if the line of sight intersects inactive regions of the beam.
In contrast, gamma-ray emission is thought to originate in the magnetospheric current sheet beyond the light cylinder rather than from the polar cap region (see~\citet{2022ARA&A..60..495P} for a review). 
This scenario is supported by global particle-in-cell plasma simulations of pulsar magnetospheres (e.g., \citealt{2014ApJ...785L..33P, 2018ApJ...855...94P, 2020A&A...642A.204C}), which have also successfully reproduced several key features observed in gamma-ray pulsars~\citep{2025A&A...695A..93C}.
Joint analyses and simulations incorporating both radio and gamma-ray observations from a large sample of pulsars can offer deeper insights into the location and nature of emission regions. Our detection of PSR J1813$-$1246 demonstrates FAST’s capability to reveal faint radio emissions from gamma-ray pulsars. 
Continued observations with FAST will increase the sample of detected sources and enable high-quality radio polarization profile measurements for many gamma-ray pulsars, providing essential data for comprehensive multiwavelength studies of pulsar magnetospheres.

\section*{Acknowledgments}

This is work is supported by the National Natural Science Foundation of China (No. 12288102, No. 12203092, No. 12041304, No. 12203045),  the Major Science and Technology Program of Xinjiang Uygur Autonomous Region (No. 2022A03013-3), the National SKA Program of China (No. 2020SKA0120100), the National Key Research and Development Program of China (No. 2022YFC2205202, No. 2021YFC2203502), the Natural Science Foundation of Xinjiang Uygur Autonomous Region (No. 2022D01B71), the Tianshan Talent Training Program for Young Elite Scientists (No. 2023TSYCQNTJ0024), the Special Research Assistant Program of CAS.  
This work made use of the data from the Five-hundred-meter Aperture Spherical radio Telescope, which is a Chinese national megascience facility, operated by National Astronomical Observatories, Chinese Academy of Sciences. The research is partly supported by the Operation, Maintenance and Upgrading Fund for Astronomical Telescopes and Facility Instruments, budgeted from the Ministry of Finance of China (MOF) and administrated by the Chinese Academy of Sciences (CAS).

\software{PRESTO~\citep{2001PhDT.......123R,2002AJ....124.1788R}, PSRCHIVE~\citep{2004PASA...21..302H}, DSPSR \citep{Straten2011}, TEMPO2~\citep{2006MNRAS.369..655H}}

\bibliography{sample63}{}
\bibliographystyle{aasjournal}

\end{document}